\documentclass{emulateapj}
\usepackage{subfigure}

\newcommand{\bmath}[1]{\mbox{\boldmath{$#1$}}}

\newcommand{\at}{{\it Athena}}

\newcommand{\del}{{\bf \nabla}}
\newcommand{\alf}{{\rm Alfv\acute{e}n}}

\newcommand{\cs}{c_{\rm s}}

\newcommand{\two}{{\rm Ideal-Lx2Ly4Lz8}}
\newcommand{\four}{{\rm Ideal-Lx4Ly8Lz8}}
\newcommand{\eight}{{\rm Ideal-Lx8Ly16Lz8}}
\newcommand{\sixteen}{{\rm Ideal-Lx16Ly32Lz8}}
\newcommand{\fourau}{{\rm Resistive-4AU}}
\newcommand{\tenau}{{\rm Resistive-10AU}}
\newcommand{\fiftyau}{{\rm Resistive-50AU}}
\newcommand{\hydrola}{{\rm Hydro-LA}}
\newcommand{\hydroha}{{\rm Hydro-HA}}

\begin{document}

\title{Turbulent Linewidths in Protoplanetary Disks:\\ Predictions from Numerical Simulations}

\author{Jacob B. Simon, Philip J. Armitage\altaffilmark{1}, and Kris Beckwith}
\affil{JILA, University of Colorado and NIST, 440 UCB, Boulder, CO 80309-0440}
\email{jbsimon@jila.colorado.edu}

\begin{abstract}
Sub-mm observations of protoplanetary disks now approach the acuity needed to measure the turbulent broadening of molecular lines. These measurements constrain disk angular momentum transport, and furnish evidence of the turbulent environment within which planetesimal formation takes place. We use local magnetohydrodynamic (MHD) simulations of the magnetorotational instability (MRI) to predict the distribution of turbulent velocities in low mass protoplanetary disks, as a function of radius and height above the mid-plane. We model both ideal MHD disks, and disks in which Ohmic dissipation results in a dead zone of suppressed turbulence near the mid-plane. Under ideal conditions, the disk mid-plane is characterized by a velocity distribution that peaks near $v \simeq 0.1 c_s$ (where $c_s$ is the local sound speed), while supersonic velocities are reached at $z > 3 H$ (where $H$ is the pressure scale height). Residual velocities of $v \approx 10^{-2} c_s$ persist near the mid-plane in dead zones, while the surface layers remain active. Anisotropic variation of the linewidth with disk inclination is modest. We compare our MHD results to hydrodynamic simulations in which large-scale forcing is used to initiate similar turbulent velocities. We show that the qualitative trend of increasing $v$ with height, seen in the MHD case, persists for forced turbulence and is likely a generic property of disk turbulence. Percent level determinations of $v$ at different heights within the disk, or spatially resolved observations that probe the inner disk containing the dead zone region, are therefore needed to test whether the MRI is responsible for protoplanetary disk turbulence.
\end{abstract} 


\altaffiltext{1}{Department of Astrophysical and Planetary Sciences, University of Colorado, Boulder}

\section{Introduction} 
Understanding the structure of 
protoplanetary disks is central to modeling the phenomenology of 
Young Stellar Objects \citep{williams11} and to theoretical studies of all phases of the formation of planetary systems. 
For a significant fraction of their lives, gas within protoplanetary disks 
is observed to actively accrete onto the central star, probably as a consequence of turbulent 
transport of angular momentum \citep{shakura73}. Globally, this accretion and redistribution 
of angular momentum results in evolution of the surface density profile \citep{lyndenbell74}, 
which limits the time scale for massive planet formation and affects quantities such as the rate of 
planet migration \citep{lubow10}. 
On smaller scales, turbulence sets the environment 
for planetesimal formation \citep{chiang10} by determining both the local concentration and 
collision velocities \citep{volk80,ormel07} of small particles that are aerodynamically coupled to the 
gas. 

Although several physical processes --- including self-gravity and the magnetorotational 
instability \cite[MRI;][]{balbus98} --- may initiate disk turbulence \citep[for a review, see][]{armitage11}, 
existing theoretical studies have 
largely been untroubled by observational validation. The most widely accepted 
constraint on disk turbulence comes from measurements of disk lifetimes \citep{haisch01} and accretion rates 
\citep{hartmann98}, which 
imply that protoplanetary disks around low-mass stars evolve and are dispersed 
on Myr time scales. This observation pins down the angular momentum transport 
efficiency if the evolution results from turbulence; the efficiency is conventionally expressed 
in terms of a \cite{shakura73} $\alpha \approx 10^{-2}$. Generically, this level of stress 
within a fluid disk implies characteristic velocity perturbations $v \sim \alpha^{1/2} c_s 
\sim 0.1 c_s$ \citep[where $c_s$ is the sound speed, e.g.][]{balbus98}, but this 
estimate is so crude as to be useful mainly for motivating further 
observations. Neither it, nor other constraints on $\alpha$ from detailed modeling of 
individual systems \citep{hueso05} provide any information on the 
nature of turbulence or on any dependence of its properties on height above the mid-plane. 

Direct determination of the strength of protoplanetary disk turbulence is 
possible by detecting the turbulent broadening of molecular lines 
observed in the infrared \citep{carr04} or sub-mm \citep{hughes11}. 
Subsonic turbulent broadening is a challenging quantity to measure, as protoplanetary disks are comprised of supersonically orbiting gas; thus,
precise measurements are needed to separate the small turbulent 
component from the dominant bulk rotation. Furthermore, in the inner disk, observed lines 
from the disk may be contaminated by outflow components \citep{bast11}. 
Nonetheless, current observations of the outer regions of disks already attain precisions comparable 
to the level ($v \sim 0.1 c_s$) where a signal can plausibly be 
expected. Using the Submillimeter Array ({\em SMA}) to observe the 
CO(3-2) transition, \citet{hughes11} derived constraints on the turbulent linewidth 
in the atmosphere of the disks surrounding the T~Tauri star TW~Hya and the 
Herbig~Ae star HD~163296\footnote{These are not ``typical" sources. TW~Hya 
is a nearby system with a favorable near face-on geometry, while HD~163296 
has a very large (500~AU) disk.}. For TW~Hya they placed an upper limit 
to the turbulent velocity of $v < 0.1 c_s$, while for 
HD~163296 they obtained a tentative detection of turbulent broadening 
corresponding to $v \approx 0.4 c_s$. Although still preliminary, 
these observations provide a clear indication that {\em ALMA}, with 
superior sensitivity and spatial resolution, will constrain 
disk turbulence to theoretically interesting levels for these and other sources. 

In this paper, our goal is to quantify the expected turbulent velocities in protoplanetary 
disks as a function of radius and height above the mid-plane. We focus on low mass 
disks (TW~Hya would be a good example) which are stable against self-gravity, 
and assume that the MRI is the sole 
source of turbulence. We compute both reference models in the ideal 
magnetohydrodynamic (MHD) limit, and physical models in which the MRI 
is partially damped by Ohmic dissipation, forming a dead zone \citep{gammie96,sano00,fromang02}. 
Particular care is taken to 
ensure that the results are numerically converged; we 
use local shearing box simulations whose convergence with spatial 
resolution has previously been demonstrated \citep{davis10,simon11}, and we explicitly 
check the effect of varying the domain size.
We also calculate purely 
hydrodynamic simulations in which turbulence is initiated through arbitrary 
large-scale forcing. By comparing these to the MHD runs, we address 
the question of whether the observable properties of disk turbulence 
can constrain the underlying mechanism that initiates angular momentum 
transport.

The plan of the paper is as follows. In \S2 we describe the numerical simulations 
in ideal MHD, non-ideal MHD, and pure hydrodynamics that form the basis of the turbulent velocity 
calculation. In \S3 we outline the velocity distribution calculation, the results 
of which are shown in \S4. \S5 discusses our results and their
implications for future observations. Finally, we summarize our conclusions
in \S 6.

\section{Simulations}
\label{simulations}

\subsection{Numerical Method}

\begin{figure}
\includegraphics*[scale=0.25,angle=0]{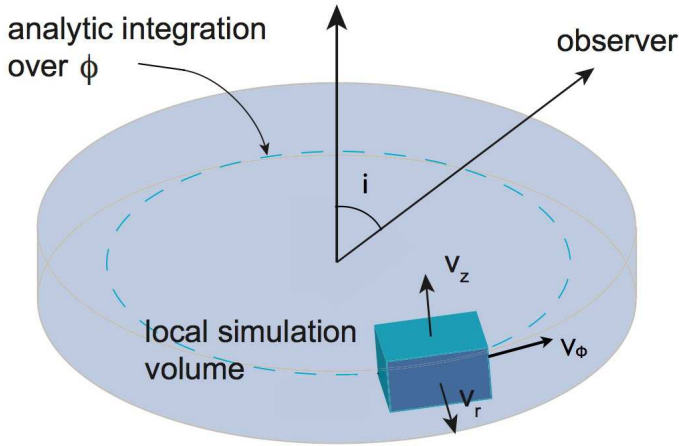}
\caption{ 
Schematic illustration of the calculation of turbulent velocity distributions and the relationship between a local simulation domain
and a disk inclined by an angle $i$ with respect to the observer. The local domain is a co-rotating patch of the larger disk, the size
of which is small enough to approximate this domain as a Cartesian box.  We extract the turbulent velocity from this local domain, appropriately
averaging over time and azimuthal angle $\phi$, as outlined in the text.
}
\label{diagram}
\end{figure}

We numerically solve the equations of magnetohydrodynamics (MHD) using
the shearing box approximation.  The shearing box is a model
for a local, co-rotating disk patch whose size is small compared to the
radial distance from the central object, $R_0$.  This allows the construction of a local
Cartesian frame $(x,y,z)$ that is defined in terms of the disk's cylindrical co-ordinates $(R,\phi,z^\prime)$ via 
$x=(R-R_0)$, $y=R_0 \phi$, and $z = z^\prime$. The local patch 
co-rotates with an angular velocity $\Omega$ corresponding to
the orbital frequency at $R_0$, the center of the box; see \cite{hawley95a} and Figure~\ref{diagram}. 
In this  frame, the equations of motion become \citep{hawley95a}:

\begin{equation}
\label{continuity_eqn}
\frac{\partial \rho}{\partial t} + \del \cdot (\rho {\bmath v}) = 0,
\end{equation}
\begin{equation}
\label{momentum_eqn}
\frac{\partial \rho {\bmath v}}{\partial t} + \del \cdot \left(\rho {\bmath v}{\bmath v} - {\bmath B}{\bmath B}\right) + \del \left(P + \frac{1}{2} B^2\right) = 2 q \rho \Omega^2 {\bmath x} - \rho \Omega^2 {\bmath z} - 2 {\bmath \Omega} \times \rho {\bmath v},
\end{equation}
\begin{equation}
\label{induction_eqn}
\frac{\partial {\bmath B}}{\partial t} - \del \times \left({\bmath v} \times {\bmath B}\right) =  -\del \times \left(\eta \del \times {\bmath B}\right).
\end{equation} 

\noindent 
where $\rho$ is the mass density, $\rho {\bmath v}$ is the momentum
density, ${\bmath B}$ is the magnetic field, $P$ is the gas pressure,
and $q$ is the shear parameter, defined as $q = -d$ln$\Omega/d$ln$R$.
We use $q = 3/2$, appropriate for a Keplerian disk.  We
assume an isothermal equation of state $P = \rho \cs^2$, where $\cs$
is the isothermal sound speed.  From left to right, the source terms
in equation~(\ref{momentum_eqn}) correspond to radial tidal forces
(gravity and centrifugal), vertical gravity, and the Coriolis force. The source term in equation~(\ref{induction_eqn}) is the
effect of Ohmic resistivity, $\eta$, on the magnetic field evolution.
Note that our system of units has the magnetic permeability $\mu = 1$.

Adopting this shearing box approximation allows for better resolution of small 
scales within the disk, at the expense of excluding global effects (those of 
scale $\sim R_0$) which could be physically important \citep{sorathia11}. 
For our purposes this trade-off is worthwhile, because we need to 
numerically resolve non-ideal MHD terms, such as Ohmic dissipation, that play an important role in the structure and evolution of
protoplanetary disks \cite[e.g.,][]{simon11}.

Our simulations use $\at$, a second-order accurate Godunov
flux-conservative code for solving the equations of MHD \cite[]{gardiner05a,gardiner08,stone08,stone10}.  
The numerical integration of the shearing box equations require additions to the \textit{Athena} algorithm, the 
details of which can be found in \cite{stone10} and the Appendix of \cite{simon11}.  Briefly, we utilize
Crank-Nicholson differencing to conserve epicyclic motion exactly and orbital advection to subtract 
off the background shear flow \cite[]{stone10}.  The $y$ boundary conditions are strictly periodic, whereas
the $x$ boundaries are shearing periodic \cite[]{hawley95a,simon11}. The vertical boundaries are the
outflow boundary conditions described in \cite{simon11}.  Finally, for simulations that include Ohmic resistivity,
the resistive term is added via first-order in time operator splitting.   We also run two simulations in the purely
hydrodynamic limit (i.e., with no magnetic fields), for which we use the HLLC Riemann solver \cite[]{toro99,stone08} appropriate
for non-MHD fluids.

\subsection{Runs, Parameters, and Initial Conditions}

Most of our calculations include MHD, and focus on the turbulent state of the MRI. The MHD 
simulations are broken down into two groups.

The first set focuses on the ideal MHD limit, in which no physical dissipation is included.
These simulations are vertically stratified, with an initial density
corresponding to isothermal hydrostatic equilibrium,

\begin{equation}
\label{density_init}
\rho(x,y,z) = \rho_o {\rm exp}\left(-\frac{z^2}{H^2}\right),
\end{equation}

\noindent where $\rho_o = 1$ is the mid-plane density, and $H$
is the scale height in the disk,

\begin{equation}
\label{scale_height}
H = \frac{\sqrt{2} \cs}{\Omega}.
\end{equation}

\noindent The isothermal sound speed, $\cs = 7.07 \times 10^{-4}$,
corresponding to an initial value for the gas pressure of $P_o = 5
\times 10^{-7}$.  With $\Omega = 0.001$, the value for the scale height
is $H = 1$.

For all ideal MHD runs except for the largest domain calculation, the initial magnetic field configuration is the
twisted azimuthal flux tube of \cite{hirose06}, with minor modifications
to the dimensions and  the value of the gas to magnetic pressure ratio, $\beta = 2P/B^2$.
In particular, the initial toroidal field, $B_y$, is given by
\begin{equation}
\label{toroidal}
B_y = \left\{ \begin{array}{ll}
\sqrt{\frac{2 P_o}{\beta_y} - \left(B_x^2+B_z^2\right)}& \quad
\mbox{if $B_x^2 + B_z^2 \neq 0$} \\
0 & \quad 
\mbox{if $B_x^2 + B_z^2 = 0$}
\end{array} \right.
\end{equation}

\noindent with the toroidal $\beta$ value $\beta_y = 100$.
The poloidal field components, $B_x$ and $B_z$,
are calculated from the $y$ component of the vector potential,

\begin{equation}
\label{vector_potential}
A_y = \left\{ \begin{array}{ll}
- \sqrt{\frac{2 P_o}{\beta_p}}\frac{a}{\pi} \left[1 + {\rm cos}\left(\frac{\pi r}{a}\right)\right] & \quad
\mbox{if $r < a$} \\
0 & \quad 
\mbox{if $r \geq a$}
\end{array} \right.
\end{equation}
\noindent where $r = \sqrt{x^2 + z^2}$ and $\beta_p = 1600$ is the
poloidal field $\beta$ value.  We choose $a$ to always be one fourth of the radial domain size; $a = L_x/4$.

The largest domain run is initialized with a volume-filling toroidal field at a constant $\beta$.
We seed the MRI in these runs by introducing random
perturbations to the density and velocity components.\footnote{These initial conditions are
identical to those of the ideal MHD simulations of \cite{simon11}.}
In order to classify any dependence of our
results on the size of the local region we examine, we have run several domain sizes:  $(L_x, L_y, L_z) = 2H\times4H\times8H$, 
$4H\times8H\times8H$, $8H\times16H\times8H$, and $16H\times32H\times8H$.   The
largest domain run has a resolution of 36 grid zones per $H$, and the resolution of each of the other runs is 32 zones per $H$.  

Ideal MHD is not a good approximation for most radii in protoplanetary disks. We have therefore run a second set 
of MHD simulations that include a height-dependent Ohmic resistivity $\eta(z)$, whose effect is to damp the MRI 
in regions where the resistivity is sufficiently high \citep{fleming00}. The first principles calculation of $\eta(z)$ at 
different radial locations within the disk is difficult, because the resistivity depends on both the sources of 
ionization and on the recombination rate. The latter is particularly uncertain, because it is tied to the unknown 
size distribution of small dust grains \citep{armitage11}. Here, we adopt a simple approach that follows that 
used previously by \cite{fleming03} and \cite{turner08}. We adopt a 
minimum mass solar nebula model \cite[]{hayashi81}, and account for ionization from X-rays, 
cosmic rays, and the radioactive decay of  ${}^{26}{\rm Al}$. For recombination, we consider only gas 
phase processes, and neglect dust physics. The resistivity is related to the electron fraction $x_e$ by,

\begin{equation}
\label{resistivity}
\eta = 6.5\times10^3 x_e^{-1} {\rm cm}^2 \rm s^{-1}
\end{equation}
\cite[]{hayashi81}, where, assuming charge neutrality, 
\begin{equation}
\label{xe}
x_e = \left(\frac{\xi}{\Gamma n_{\rm H}}\right)^{1/2}.
\end{equation}
Here $\xi$ is the ionization rate, comprised of the cosmic ray ionization rate,
\begin{equation}
\label{cr_rate}
\xi_{\rm CR} = 10^{-17} \left(e^{-\Sigma_a(z)/100 {\rm g}{\rm cm}^{-2}}+e^{-\Sigma_b(z)/100 {\rm g}{\rm cm}^{-2}}\right) \rm s^{-1}
\end{equation}
the X-ray ionization rate \cite[]{turner08},
\begin{equation}
\label{xr_rate}
\xi_{\rm XR} = 2.6\times10^{-15} \left(\frac{r}{\rm AU}\right)^{-2} \left(e^{-\Sigma_a(z)/8 {\rm g}{\rm cm}^{-2}}+e^{-\Sigma_b(z)/8 {\rm g}{\rm cm}^{-2}}\right) \rm s^{-1}
\end{equation}
and the ${}^{26}{\rm Al}$ decay rate, which is constant at $4\times10^{-19}\rm s^{-1}$ \cite[]{stepinski92}.  In these 
expressions $\Sigma_a(z)$ and $\Sigma_b(z)$ are the column density
lying above and below a vertical point $z$.
The recombination rate, $\Gamma$, is
\begin{equation}
\label{recomb_rate}
\Gamma = 8.7\times10^{-6} \left(\frac{T}{\rm K}\right)^{-1/2} {\rm cm}^3 \rm s^{-1},
\end{equation}
\cite[]{glassgold86}, and $n_{\rm H}$ is the number density of hydrogen,
\begin{equation}
\label{nh}
n_{\rm H} = 5.8\times10^{14} \left(\frac{r}{\rm AU}\right)^{-11/4} e^{-z^2/H^2}, 
\end{equation}
which is proportional to the gas density, $\rho$ \citep{wardle07}.

We have run three shearing boxes with this height-dependent resistivity, each of which
was restarted from the $4H \times 8H \times 8H$ ideal simulation at 100 orbits into the
integration but with the appropriate resistivity profile added.  The first, centered on $R_0 = 4 {\rm AU}$ has a large
dead zone within $\sim 2 H$ of the mid-plane surrounded by two MRI active regions.  
The second region is centered on $R_0 = 10 {\rm AU}$, which
is an intermediate region where the resistivity near the mid-plane is large enough to cause some damping of MRI turbulence but not sufficiently large to completely
quench the turbulence, resulting in episodic bursts of mid-plane turbulence resembling the constant resistivity runs of \cite{simon11}.
This dramatic variability results from the competition between Ohmic damping of MRI turbulence and the shearing of residual radial field into toroidal field of sufficient strength to reactivate
the turbulence.  Finally, the third shearing box is centered on $R_0 = 50 {\rm AU}$ and has sustained turbulence throughout the domain as the
resistivity is not large enough to damp out the MRI.

Thus, we have three different physical regimes for MRI-driven turbulence: one that resembles the classic layered accretion model \citep{gammie96}, one that is relatively close to ideal MHD, and one intermediate regime that leads to large amplitude variability in turbulence levels.
We should note that the radial locations of these regimes are subject to some uncertainty given the particular disk model that we have adopted.
 If we adopted another model, such as a constant $\alpha$ disk model for example, then we may find the  radial locations of our three regimes 
 would be different.  

\begin{widetext}
\begin{deluxetable}{l|ccccccccc}
\tabletypesize{\small}
\tablewidth{0pc}
\tablecaption{Shearing Box Simulations\label{tbl:sims}}
\tablehead{
\colhead{Label}&
\colhead{Domain Size}&
\colhead{Resistivity?}&
\colhead{MHD/Hydro} &
\colhead{KE}&
\colhead{$(|v_h|/\cs)_{\rm peak}$}&
\colhead{$(|v_z|/\cs)_{\rm peak}$}&
\colhead{$\%~|v_h|/\cs > 1$}&
\colhead{$\%~|v_z|/\cs > 1$} \\
\colhead{ }&
\colhead{($L_x \times L_y \times L_z) H$}&
\colhead{ }&
\colhead{ }&
\colhead{ }&
\colhead{for $z > 3H$}&
\colhead{for $z > 3H$}&
\colhead{for $z > 3H$}&
\colhead{for $z > 3H$}    } 
\startdata
\two & $2 \times 4 \times 8$ & No & MHD & 0.03 & 0.38 & 0.37 & 3.2 & 1.9 \\
\four & $4 \times 8 \times 8$ & No &  MHD & 0.02 & 0.54 & 0.45 & 5.1 & 5.3 \\
\eight & $8 \times 16 \times 8$ & No &  MHD & 0.02 & 0.61 & 0.48 & 9.3 & 7.5 \\
\sixteen & $16 \times 32 \times 8$ & No &  MHD & 0.03 & 0.66 & 0.54 & 13.6 & 9.0 \\
\fourau & $4 \times 8 \times 8$ & $\eta(z)$ at 4 AU & MHD & 0.002 & 0.56 & 0.48 & 9.5 & 10.5 \\
\tenau &$4 \times 8 \times 8$ & $\eta(z)$ at 10 AU & MHD & 0.02 & 0.56\tablenotemark{a}, 0.54\tablenotemark{b} & 0.45\tablenotemark{a}, 0.52\tablenotemark{b} & 12.1\tablenotemark{a}, 9.1\tablenotemark{b} & 12.7\tablenotemark{a}, 8.1\tablenotemark{b} \\
\fiftyau & $4 \times 8 \times 8$ & $\eta(z)$ at 50 AU & MHD & 0.03 & 0.54 & 0.48 & 6.0 & 7.7 \\
\hydroha & $4 \times 8 \times 8$ & No & Hydro & 0.07 & 0.45 & 0.54 & 6.7 & 8.3 \\
\hydrola & $4 \times 8 \times 8$ & No & Hydro & 0.007 & 0.22 & 0.18 & 0.03 & 0.08 \\
\enddata
\tablenotetext{a}{~Low stress state value}\\
\tablenotetext{b}{~High stress state value}
\end{deluxetable}
\end{widetext}

Finally, as one of our goals in this work is to explore how sensitive our derived turbulent velocity distributions are
to the underlying mechanism for generating turbulence, we have also run forced turbulence hydrodynamic shearing boxes. These
runs are also isothermal, vertically stratified with an initially exponential density profile (Equation~\ref{density_init}), and have the same values for $\rho_o$, $\cs$, and $\Omega$. In these cases,
we do not evolve the induction equation (${\bmath B} = 0$), and we instead add a force to the momentum equation,

\begin{eqnarray}
\label{forcing}
{\bmath f(x,y,z)} & = & \rho A [ {\rm sin}(k_x x) {\rm cos}(k_y y) {\rm cos}(k_z z)\hat{\bmath x} - {\rm cos}(k_x x) {\rm sin}(k_y y)  \nonumber \\
& & \times {\rm cos}(k_z z)\hat{\bmath y}+{\rm sin}(k_z z)\hat{\bmath z}]
\end{eqnarray}

\noindent
where $k_x = 4\pi/L_x$, $k_y = 8\pi/L_y$, $k_z = 8\pi/L_z$, and $A$ is the amplitude of the forcing.  This forcing is only applied for $|z| \le 2 H$. We have produced two of these calculations, one with $A = 10^{-3}$
and one with $A = 10^{-4}$.  These calculations were performed at a resolution of 36 zones per $H$ and at a domain size of $4H\times8H\times8H$.

Evolving the MHD simulations becomes difficult if there are magnetized regions of very low density, 
where a large $\alf$ speed results in a small timestep. Moreover, errors in energy make it hard 
to evolve regions of very strong field relative to gas pressure without encountering numerical problems. To avoid 
these problems, we apply a density floor at a level of $10^{-4}$ of the initial mid-plane density
throughout the physical domain in our MHD simulations.  We also include a density floor in our hydro simulations, which we set to $10^{-8}$.  
The hydrodynamic floor can be much lower since there is no $\alf$ speed restriction on the timestep.

Table~\ref{tbl:sims} summarizes the runs, along with some basic properties of the turbulence that they 
generate. The ideal MHD runs are labelled with ``Ideal" as a prefix and then the $x,y,z$ domain size in units of $H$.  The resistive runs 
have the prefix ``Resistive" appended with the domain's radial location in our model disk.  Finally, the forced hydrodynamic runs are prefixed with ``Hydro", and suffixed with HA (for high-amplitude; $A = 10^{-3}$) or LA (for low-amplitude; $A = 10^{-4}$).

\section{Velocity Distribution Calculation}
\label{calculation}

In this work, we do not consider any radiative transfer effects or an emission model. Instead, we determine how the 
density-weighted turbulent velocity distribution
depends on location within a protoplanetary disk and on the physics that we include. Although 
not equivalent to an observed turbulent line profile,
the velocity distribution gives us the probability of observing emission at a particular velocity shift along the line of sight.  

The line-of-sight ($los$) turbulent velocity of a patch of disk will depend on the inclination
angle of the disk $i$, and the azimuthal angle $\phi$ around the disk center (see Figure~\ref{diagram}),
\begin{equation}
\label{vlos}
v_{los} = v_r {\rm cos} (\phi) {\rm sin}(i) - v_{\phi} {\rm sin} (\phi) {\rm sin} (i) + v_z {\rm cos} (i),
\end{equation}
where ($v_r, v_{\phi}, v_z$) is the turbulent velocity field in cylindrical coordinates centered on the disk.  We can rewrite this velocity field in terms of shearing box
coordinates $(x,y,z)$ as $v_x = v_r$, $v_y = v_{\phi}$, and $v_z = v_z$; this is a trivial transformation because we are interested in
the magnitude of the turbulent velocity fluctuations, which is the same in either frame.  
In principle, spatially resolved observations of disks at different inclinations could yield 
independent constraints on all three velocity components. For simplicity, we consider here 
just two components, a vertical turbulent velocity and an azimuthally averaged combination 
of $v_x$ and $v_y$ that corresponds to an average over an annulus of the disk.
This ``horizontal" (i.e., disk planar) turbulent velocity magnitude is defined as,
\begin{equation}
\label{vh}
|v_h| \equiv \frac{1}{2\pi}\int_0^{2\pi} |v_x {\rm cos}(\phi) - v_y {\rm sin}(\phi)| d\phi.
\end{equation}
Practically, we extract $v_x$, $v_y$, and $|v_z|$ from our shearing box calculations, analytically average $v_x$ and $v_y$ as described above to get
$|v_h|$, and then time-average the resulting density-weighted velocity distributions over some period during the saturated state.  This period varies greatly
between our various runs and was chosen based upon the system being in a statistically steady state and the horizontally averaged density being above the floor in the upper $|z|$ regions.  Finally, to represent different line penetration depths,
we calculate each distribution for $z > 3H$, $z > 2H$, $z > H$, and $z > 0$.

\section{Results}
\label{results}

Before discussing the velocity distributions themselves, we first explore two basic diagnostics of the turbulent flows 
that are generated in the MHD and hydrodynamic simulations.  
The first diagnostic is the turbulent kinetic energy normalized by the gas pressure,
\begin{equation}
\label{ke}
{\rm KE} \equiv \overline{\frac{1}{2}\frac{\left\langle \rho v^2 \right\rangle}{\left\langle P\right\rangle}},
\end{equation} 
where the brackets denote a volume average (over the entire domain), and the overbar denotes a time-average over a suitable interval in which the
turbulence is statistically steady. 

Table~\ref{tbl:sims} displays the normalized kinetic energy for all simulations. The time average is done onward from orbit 50 for all simulations except for Hydro-LA, in which it was done from orbit 150 onwards.  With the exception of the strong dead zone shearing box (Resistive-4AU), KE $\sim 0.02-0.03$ for all MRI simulations.  The time history of the kinetic energy in Resistive-10AU is highly variable, up to a factor of 4. Yet, the time-averaged value of this kinetic energy is consistent with the other MRI calculations.  By design, the two forced hydro simulations bracket
the MRI simulations in terms of kinetic energy.  Hydro-HA has more kinetic energy than most of the MRI simulations by a factor of $\sim 3$, whereas Hydro-LA has less kinetic energy by about the same factor.

\begin{figure}
\includegraphics*[scale=0.45,angle=0]{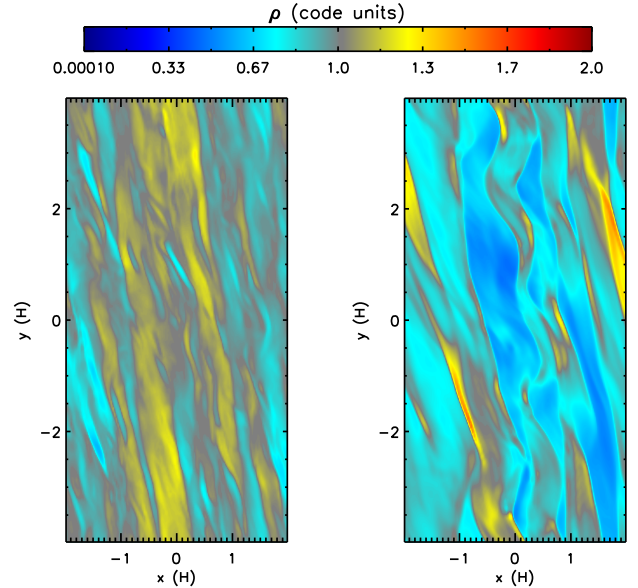}
\caption{ 
Snapshot of the mid-plane gas density at 100 orbits.  The left panel is from run Resistive-50AU and represents an MRI calculation.  The right panel
is from Hydro-HA and is a forced turbulence simulation.  Both calculations show the presence of spiral density waves.
}
\label{sdw_plot}
\end{figure}

While we have established that we can force turbulence to roughly the same kinetic energy amplitude as the MRI-driven cases, it is worth asking what the structure of the forced turbulence is.  Do the forced turbulence runs resemble their forcing functions at late times,  or does a different structure emerge? In Figure~\ref{sdw_plot}, we plot the density in the mid-plane of the Resistive-50AU run and the Hydro-HA run at 100 orbits into the evolution.  The two runs are visually nearly indistinguishable, and even in the forced turbulence case, there exist spiral density waves that propagate through the domain.  Indeed, the auto-correlation function for the gas density \citep{guan09,nelson10} returns a density structure that is very similar between the two runs.  This similarity may be a result of the choice of forcing function for the hydro calculations; if we had chosen some other forcing function, perhaps these density waves would not exist or would look different to the MRI case.  However, 
\cite{heinemann09} suggest that these waves can be generally produced by disk turbulence, not necessarily restricted to that which is MRI-driven.  In this respect, our hydro calculations are a representation of potential forms of disk turbulence that produce these waves other than the MRI. 

\begin{figure}[p]
\begin{minipage}[!ht]{8cm}
\begin{center}
\includegraphics[width=0.95\textwidth,angle=0]{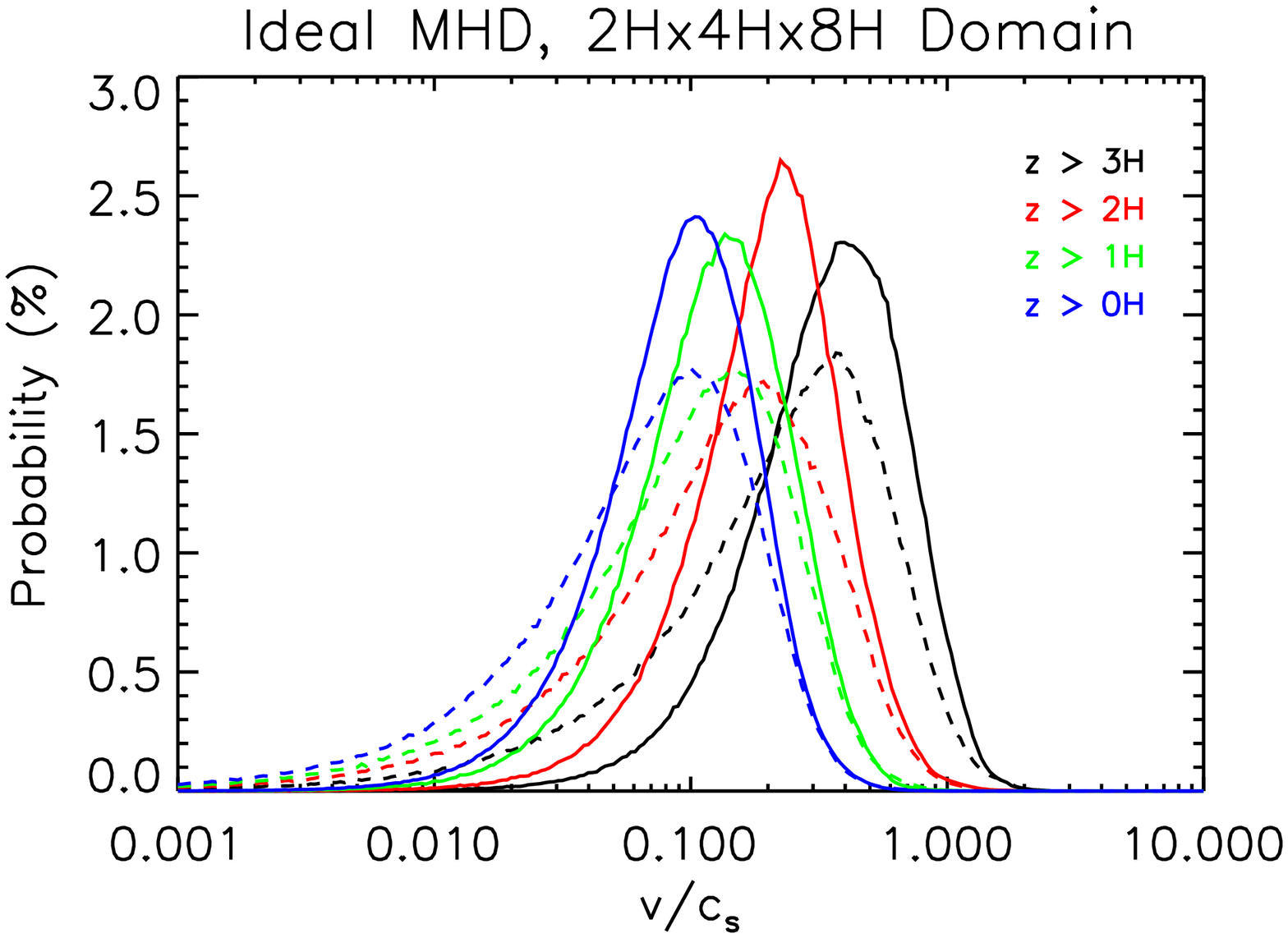}
\end{center}
\end{minipage}
\begin{minipage}[!ht]{8cm}
\begin{center}
\includegraphics[width=0.95\textwidth,angle=0]{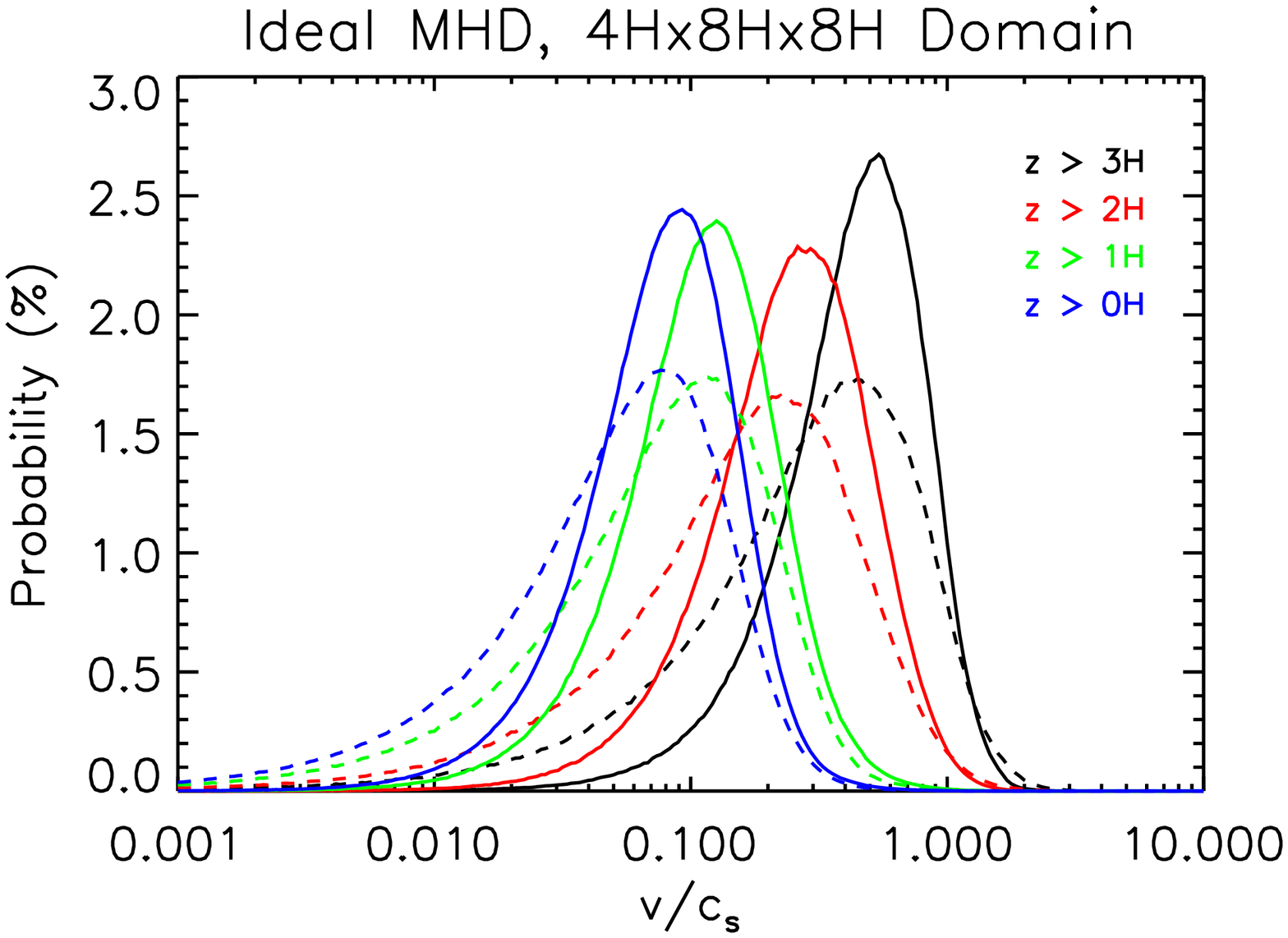}
\end{center}
\end{minipage}
\begin{minipage}[!ht]{8cm}
\begin{center}
\includegraphics[width=0.95\textwidth,angle=0]{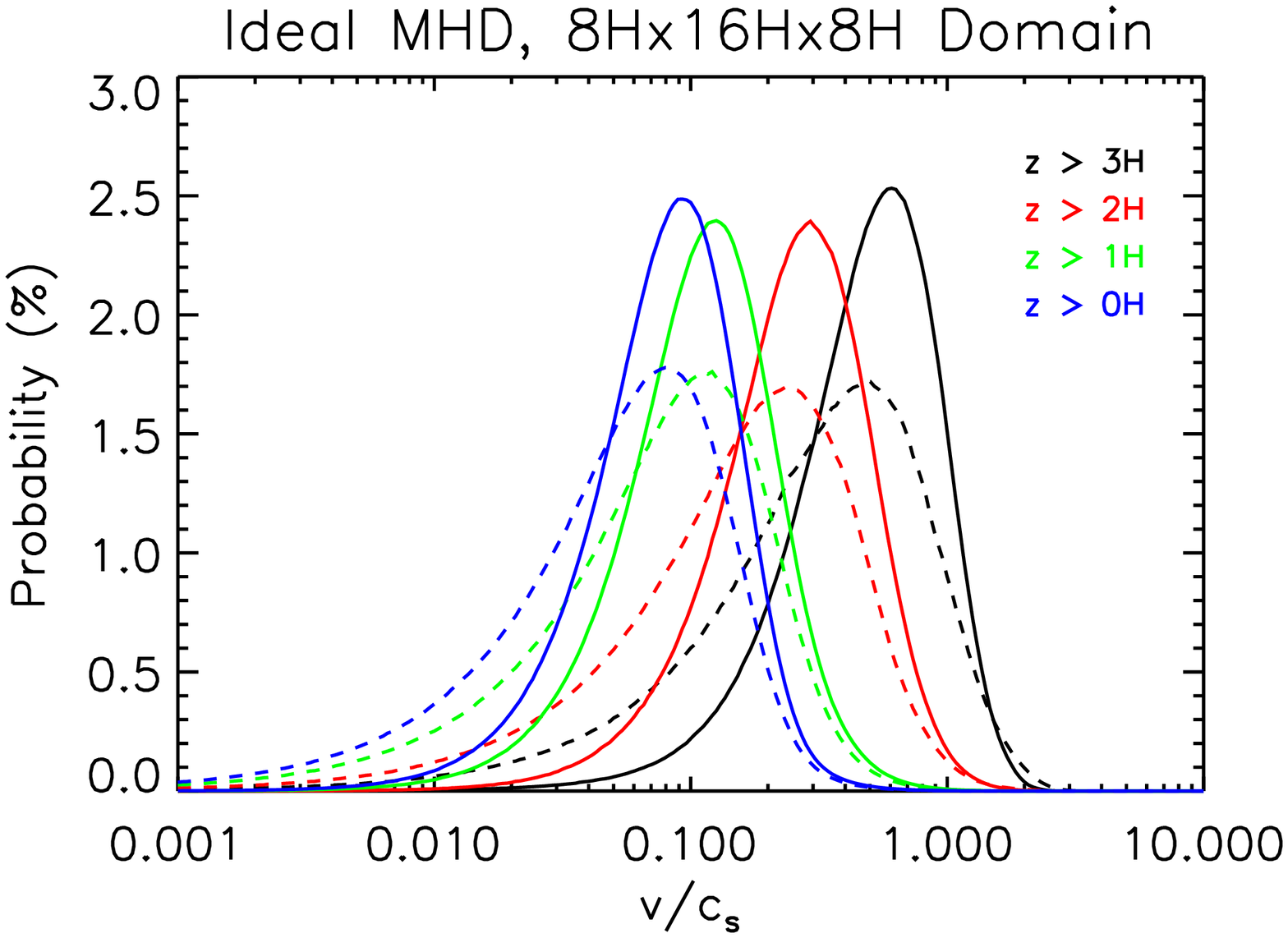}
\end{center}
\end{minipage}
\begin{minipage}[!ht]{8cm}
\begin{center}
\includegraphics[width=0.95\textwidth,angle=0]{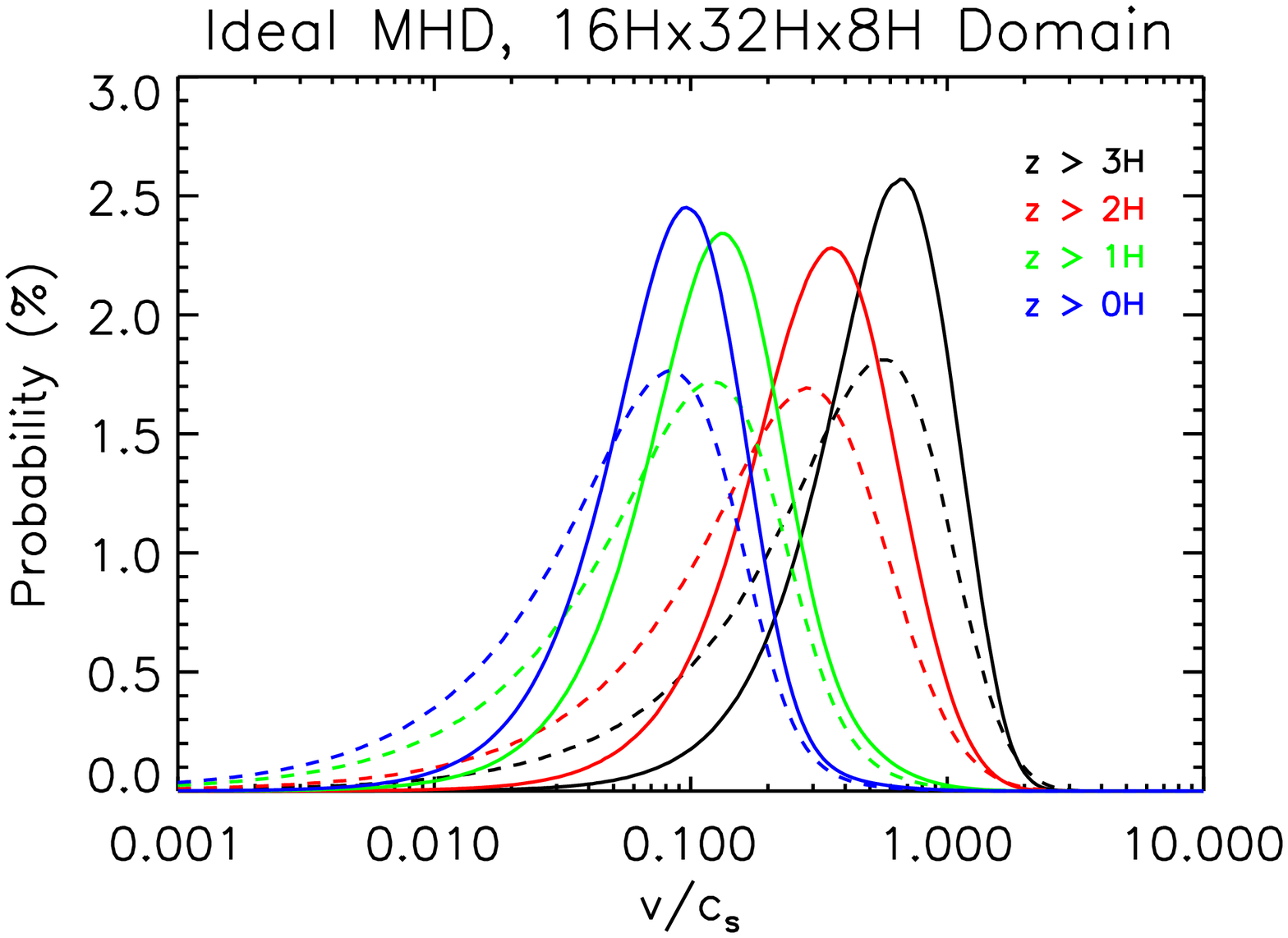}
\end{center}
\end{minipage}
\caption{
Turbulent velocity distributions in the ideal MHD calculations.  Each panel corresponds to a different local domain size, and the different colors in each
figure correspond to different depths over which the distribution is calculated, as labeled. The dashed lines are the vertical turbulent velocity $|v_z|/\cs$, and
the solid lines are the azimuthally averaged disk planar velocity $|v_h|/\cs$.  Most of the turbulent velocities are in the range $|v|/\cs \sim 0.1-1$. 
}
\label{vturb_ideal}
\end{figure}

Figure~\ref{vturb_ideal} displays the density-weighted turbulent velocity distribution for several integration depths and shearing box domain sizes, all in the
ideal MHD limit.  The dashed lines are $|v_z|$ and the solid lines are $|v_h|$.  
The most striking feature of these plots is the rapid increase in the velocity of the peak of the distribution 
as one moves higher in the disk. For $z > 0$ (upper half of the disk) the distribution peaks at about 
10\% of the sound speed, but this velocity increases to about 50\% of $c_s$ for $z > 3H$. The width of the distributions is quite large; for the $4H\times8H\times8H$ domain, there is a $\sim 90\%$ probability that $|v_h|/\cs$ lies between 0.1 and 1.  There does not appear to be a strong difference between the $|v_z|$ and $|v_h|$ distributions, with the latter being slightly more sharply peaked
and at a slightly higher $|v|/\cs$ than the former.  This suggests that the inclination of the disk will only weakly play a role in the observed turbulent velocity.
Finally, convergence of the peak velocity with domain size appears 
to have been attained for the $4H\times8H\times8H$ domain; this suggests
that all the essential physics involved in setting the magnitude of velocity fluctuations is captured by this intermediate-sized domain.   

Another interesting feature to note is the non-negligible supersonic velocity component to the distribution above $3H$.  Integrating
over the distribution for $|v|/\cs > 1$ yields roughly 10\% of the turbulent velocity being supersonic for box sizes larger than $8H\times16H\times8H$.  The smaller domains have
smaller supersonic components: $\sim 5\%$ and $\sim 1\%$ for the $4H\times8H\times8H$ and $2H\times4H\times8H$, respectively.  The origin of these supersonic
velocities is presumably the steepening of initially subsonic waves as the density decreases away from the mid-plane.  
Similar physical effects have been seen in many prior simulations of stratified disks \citep{stone96,flock11,beckwith11}.   Along with the recently studied dissipation of current sheets in disk corona \citep{hirose11}, shock heating from these
supersonic motions could potentially play an important role in the thermodynamic properties of disk atmospheres.  We will further explore these thermodynamic issues in future publications. We include the peak of the $|v|/\cs$ distribution and the percentage of the distribution with supersonic velocities for $z > 3 H$ in Table~\ref{tbl:sims}.

\begin{figure}[p]
\begin{minipage}[!ht]{8cm}
\begin{center}
\includegraphics[width=0.95\textwidth,angle=0]{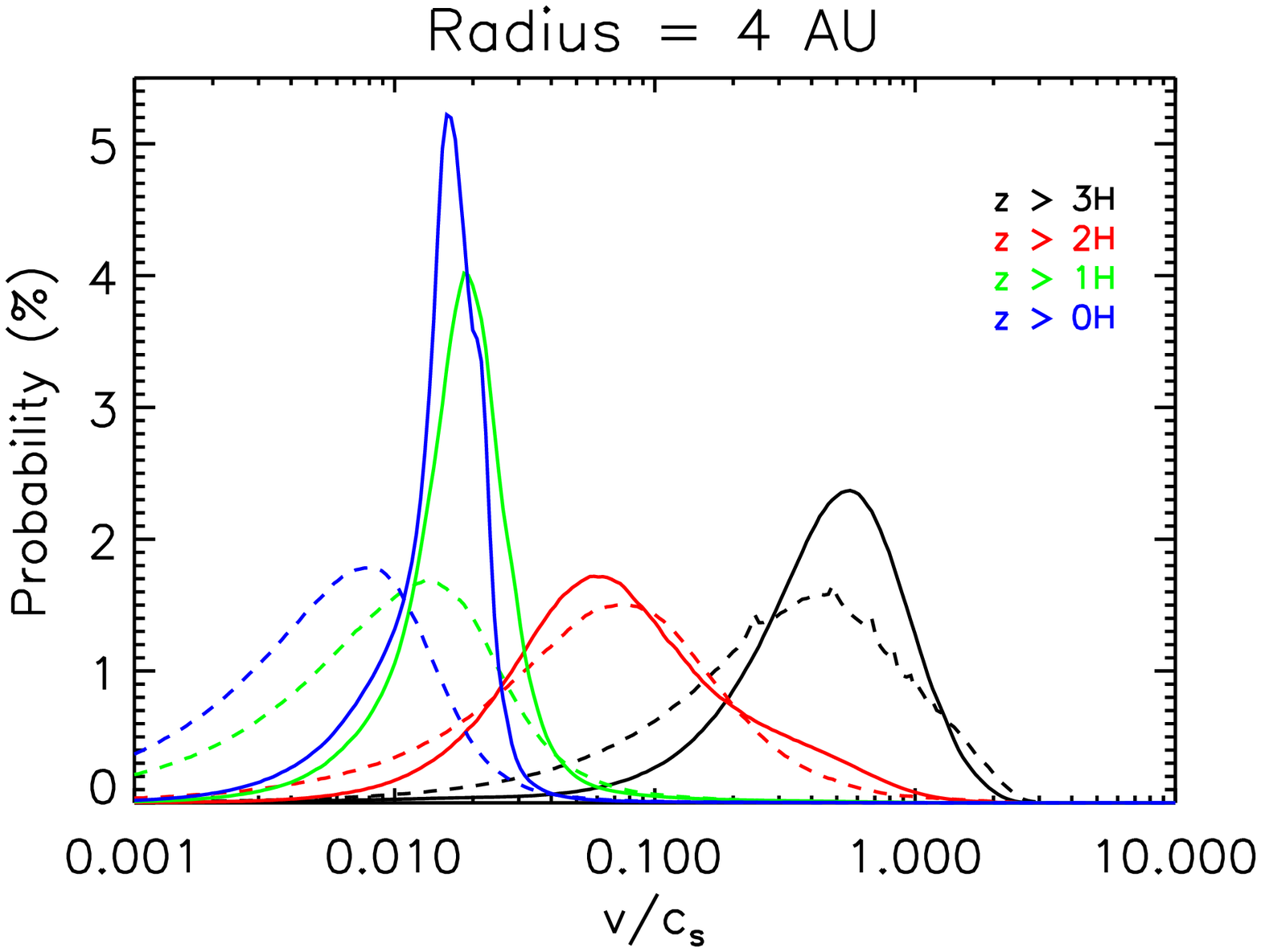}
\end{center}
\end{minipage}
\begin{minipage}[!ht]{8cm}
\begin{center}
\includegraphics[width=0.95\textwidth,angle=0]{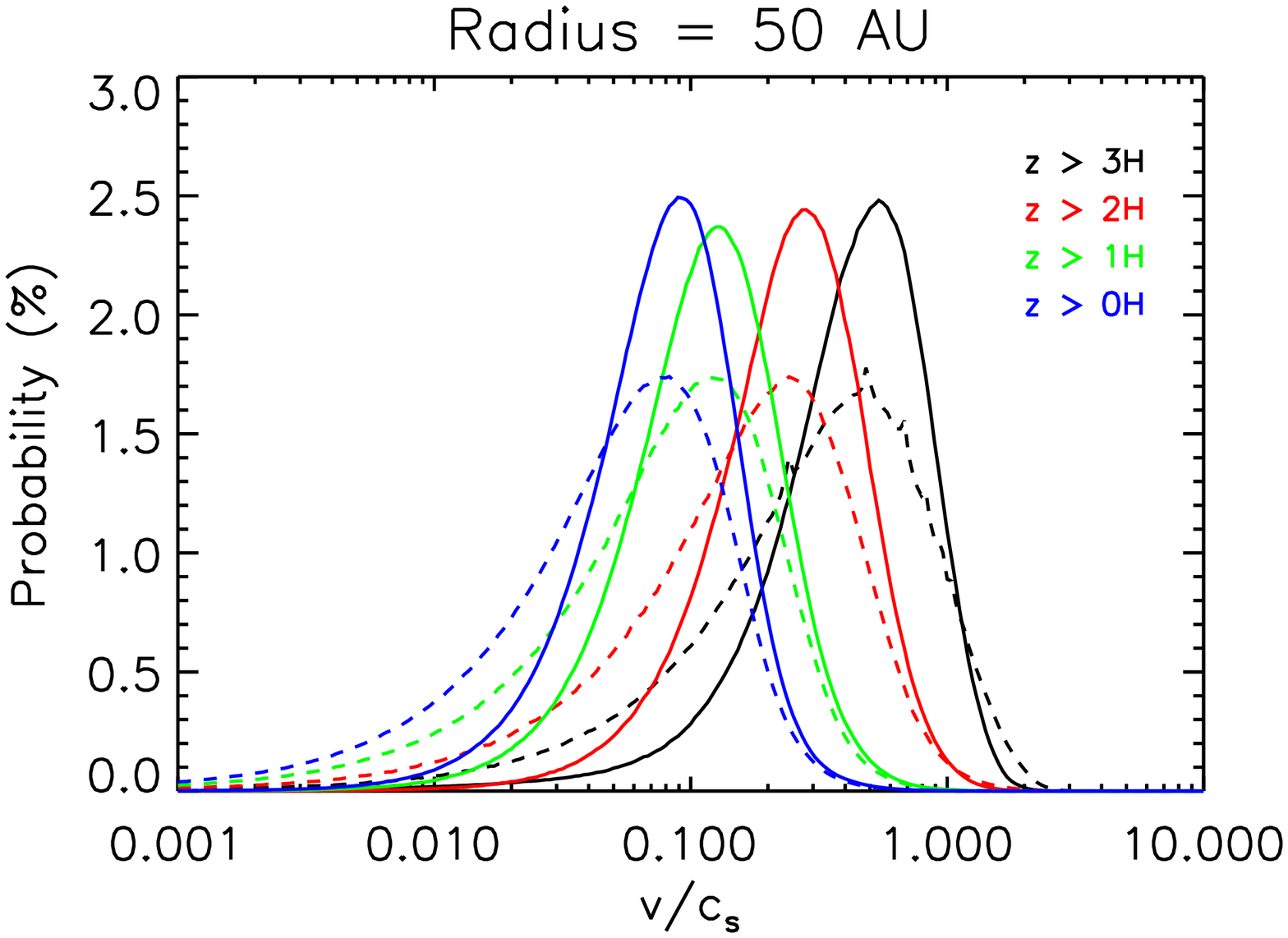}
\end{center}
\end{minipage}
\begin{minipage}[!ht]{8cm}
\begin{center}
\includegraphics[width=0.95\textwidth,angle=0]{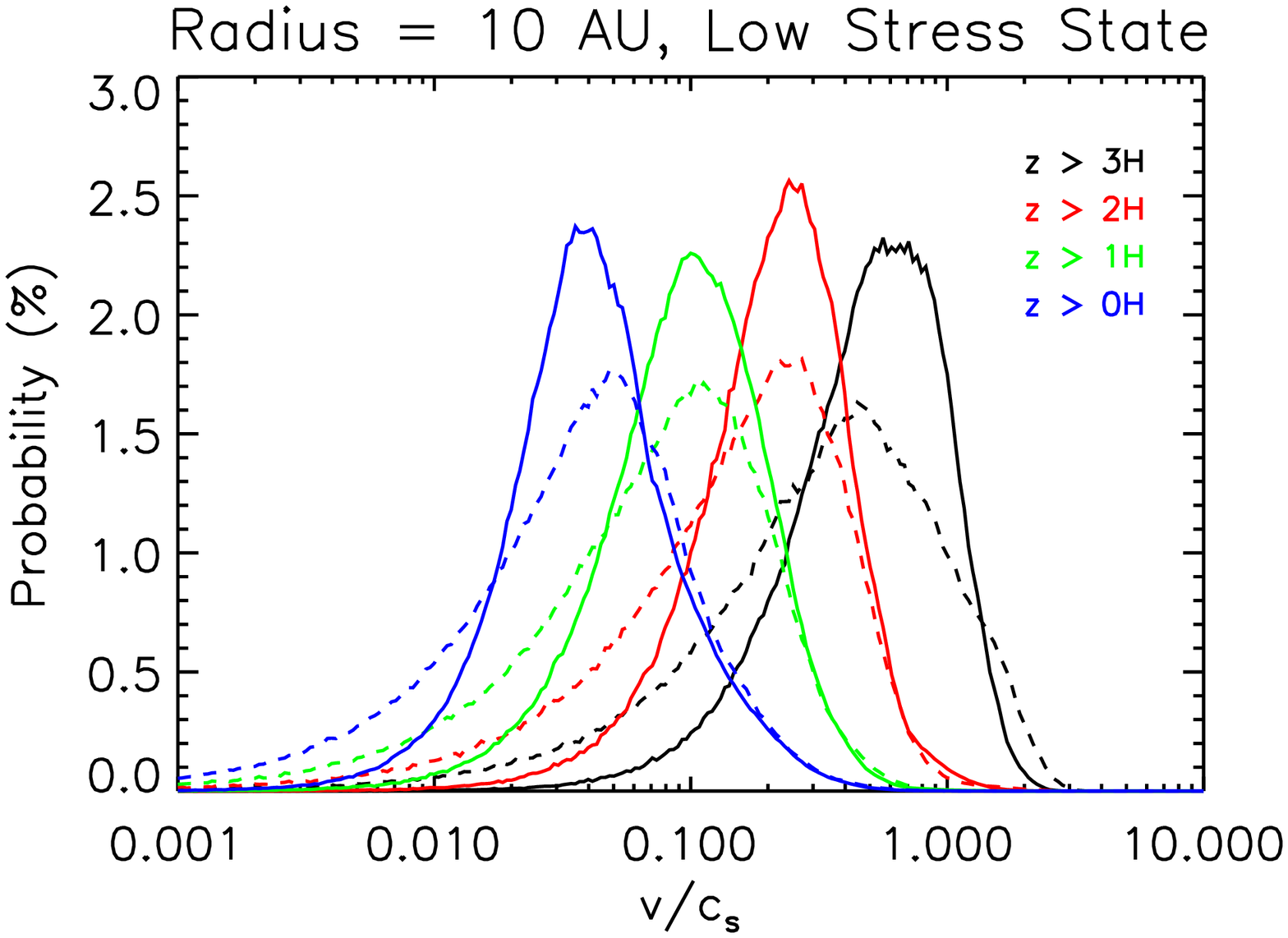}
\end{center}
\end{minipage}
\begin{minipage}[!ht]{8cm}
\begin{center}
\includegraphics[width=0.95\textwidth,angle=0]{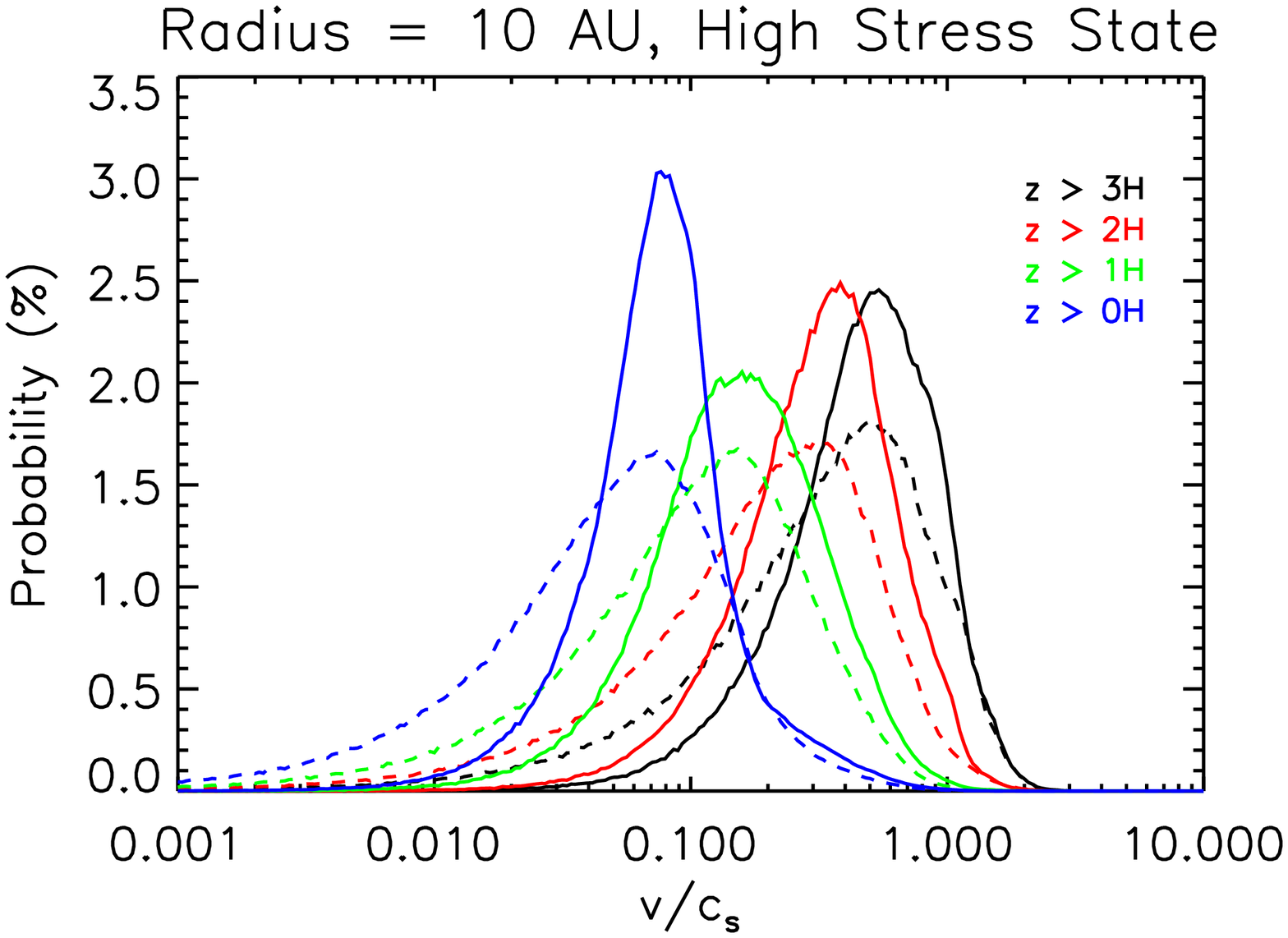}
\end{center}
\end{minipage}
\caption{
Turbulent velocity distributions for the resistive MHD calculations. The lower two panels are the box at 10 AU in a state characterized by low turbulent stress (top)
and high turbulent stress (bottom). The different colors in each
figure correspond to different depths over which the distribution is calculated, as labeled. The dashed lines are the vertical turbulent velocity $|v_z|/\cs$, and
the solid lines are the azimuthally averaged disk planar velocity $|v_h|/\cs$.  The 4 AU calculation shows the presence of a strong dead zone, as the velocity
distribution peaks at a much lower value towards the mid-plane. 
}
\label{vturb_nd}
\end{figure}

In Figure~\ref{vturb_nd} we plot the same velocity distributions for non-ideal (resistive) shearing boxes computed at 
different radial locations. Since the simulation conducted at 10~AU is highly 
variable, we show results that correspond both to the 
high stress turbulent state, and to the low stress state. Considering first the shearing box centered
on 4 AU, it appears that as one probes regions closer to the mid-plane, the turbulent velocity fluctuations drop dramatically, with a peak in the distribution
at $|v|/\cs \sim 0.02$.  This is not surprising since the dead zone in this run extends to about $\pm 2H$, and the velocity fluctuations induced
by the active layers appear within the dead zone region \cite[e.g.,][]{fleming03,simon11}.  Above $3H$ the velocity distribution is very similar to 
the ideal MHD cases, including the presence of supersonic velocities.

Moving outward to near the outer edge of the dead zone, 
the 10 AU simulation yields a velocity distribution that varies slightly, depending on whether or not
the system is in the ``high state" or the ``low state".  The low state appears to be intermediate in the velocity distribution between
the 4 AU and ideal MHD cases, whereas the high state resembles the ideal MHD distribution more closely.  In both states, the velocity distribution near the disk
surface peaks around $|v|/\cs \sim 0.5$ with a substantial supersonic tail, again agreeing with the other simulations. 
Finally, the shearing box centered on 50 AU has a distribution very similar to the ideal MHD case, consistent with the notion that the resistivity
is small enough at this radius to not significantly affect the MRI.

\begin{figure}
\begin{minipage}[!ht]{8cm}
\begin{center}
\includegraphics[width=1\textwidth,angle=0]{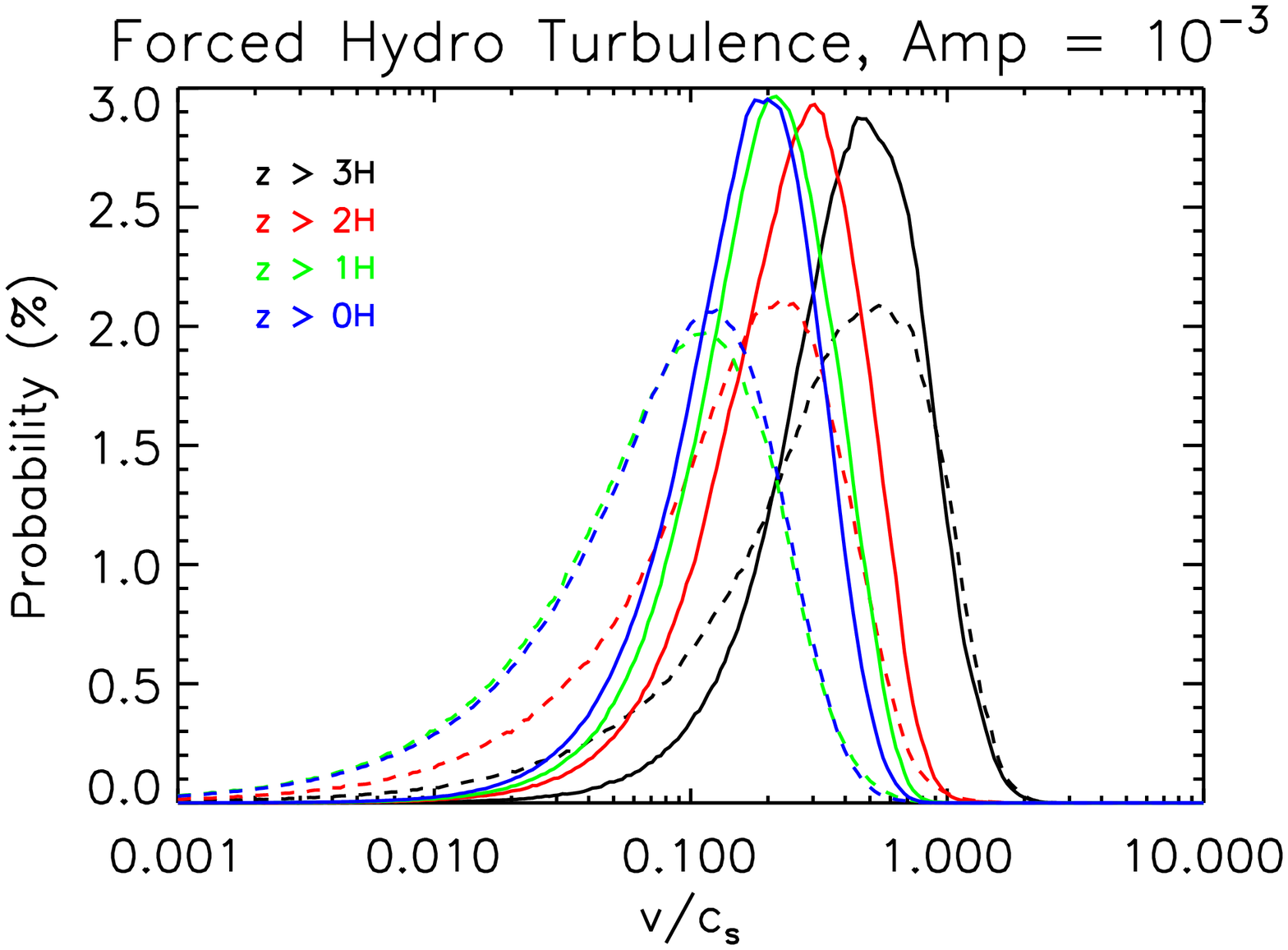}
\end{center}
\end{minipage}
\begin{minipage}[!ht]{8cm}
\begin{center}
\includegraphics[width=1\textwidth,angle=0]{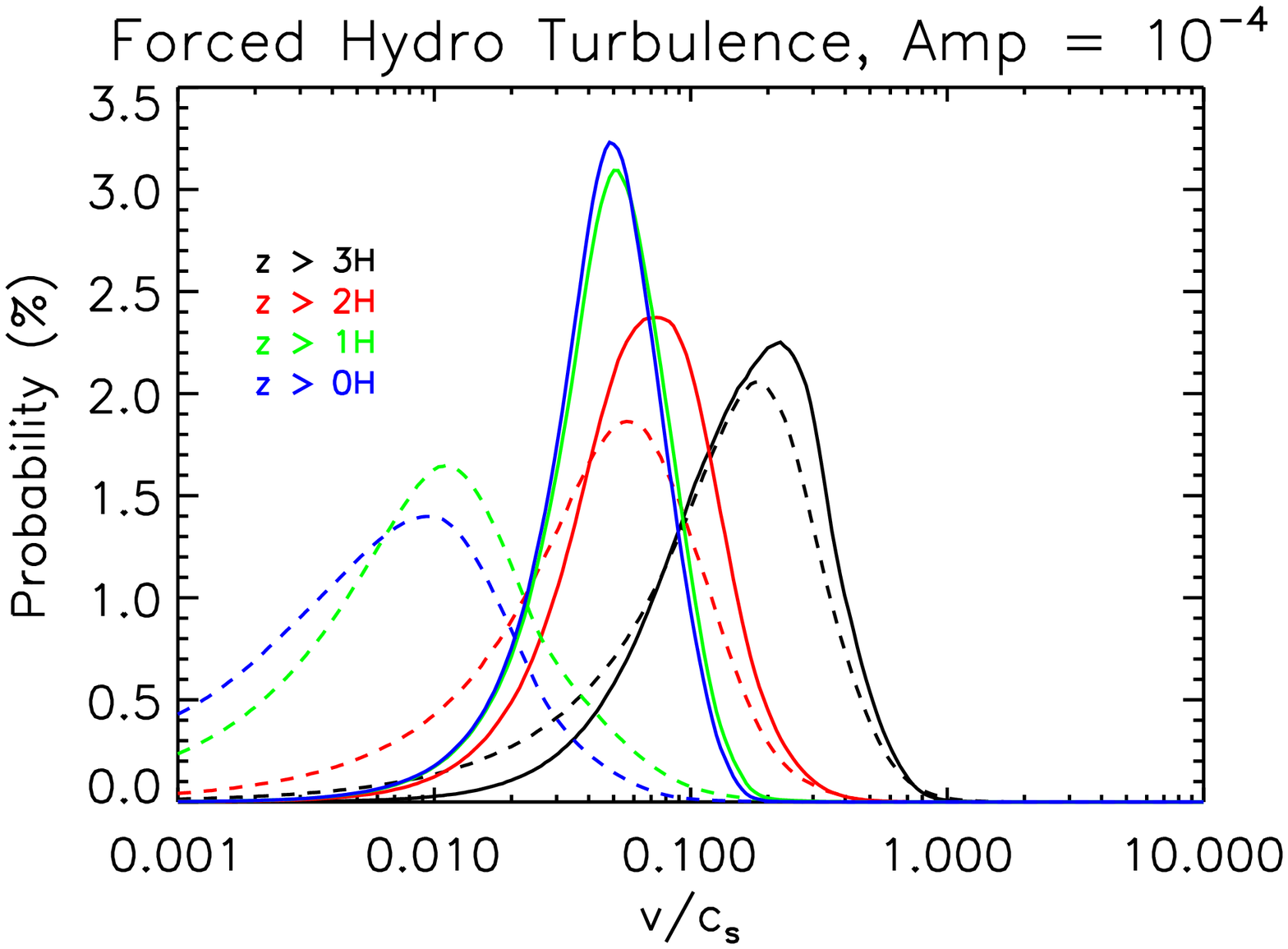}
\end{center}
\end{minipage}
\caption{
Turbulent velocity distributions for the forced hydro calculations. The top panel is the run forced with amplitude $10^{-3}$ and
the bottom panel has forcing with amplitude $10^{-4}$. The different colors in each
figure correspond to different depths over which the distribution is calculated, as labeled. The dashed lines are the vertical turbulent velocity $|v_z|/\cs$, and
the solid lines are the azimuthally averaged disk planar velocity $|v_h|/\cs$.  
}
\label{vturb_fh}
\end{figure}

Figure~\ref{vturb_fh} displays the velocity distributions for the two forced hydro runs.  Hydro-HA has a distribution quite similar to those
in the MRI calculations.   However, there is a weaker dependence of the turbulent velocity on the height from the mid-plane; the peak $|v|/\cs$
values lie between 0.2 and 0.5.   Again, the peaks of the $|v_z|/\cs$ and $|v_h|/\cs$ distributions are quite similar.  There is also a 
significant supersonic component to the $z > 3H$ distribution; $\sim$~7-8\% of the distribution has $|v|/\cs > 1$. 
The distribution peaks for Hydro-LA are lower, which is not surprising since there is less kinetic energy in this run.  However, despite an order of magnitude difference
in the saturated kinetic energies, the peak velocity for $z > 3H$ is only a factor of 2.5 lower in Hydro-LA than in Hydro-HA.  The mid-plane velocities are significantly lower
in Hydro-LA than in Hydro-HA, however.  These results suggest that even when forced with a lower amplitude, the turbulent velocities can steepen significantly in the lower density regions away from the mid-plane.  Taken together, the characteristic velocities in the forced hydro runs are not very different than those in the MRI cases.

\section{Discussion and Implications for Observations}
\label{discussion}

\begin{figure}
\begin{center}
\includegraphics[width=0.5\textwidth,angle=0]{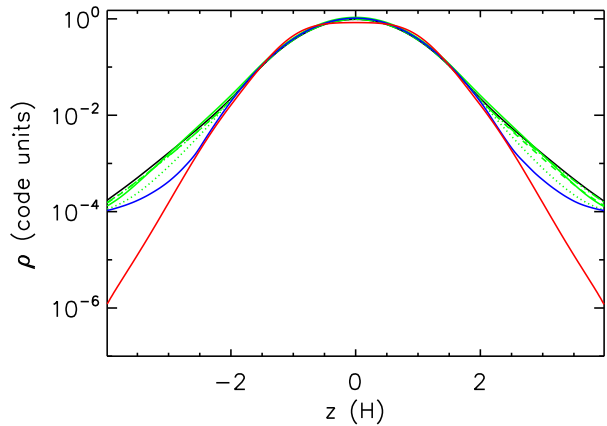}
\end{center}
\caption{
Time- and horizonally-averaged vertical density profiles for a subset of the shearing box simulations.  The time average was done from orbit 50 onward.  The density is in code units ($\rho_{z=0} \sim 1$), and $z$ is in units of $H$.  The green curves correspond to the ideal MHD simulations, and the solid green curve
is Ideal-Lx2Ly4Lz8, the dashed green curve is Ideal-Lx4Ly4Lz8, the dotted green curve is Ideal-Lx8Ly8Lz16, and the triple-dot dash green curve is Ideal-Lx16Ly32Lz8.
The blue line corresponds to Resistive-4AU, the black line is Resistive-50AU, and the red line is Hydro-HA.   The forced hydro run has a nearly Gaussian density profile, whereas the MHD calculations have a Gaussian profile for $|z| \lesssim 2H$, outside of which the density gradient flattens out.
}
\label{dz_all}
\end{figure}

\begin{figure}
\begin{center}
\includegraphics[width=0.5\textwidth,angle=0]{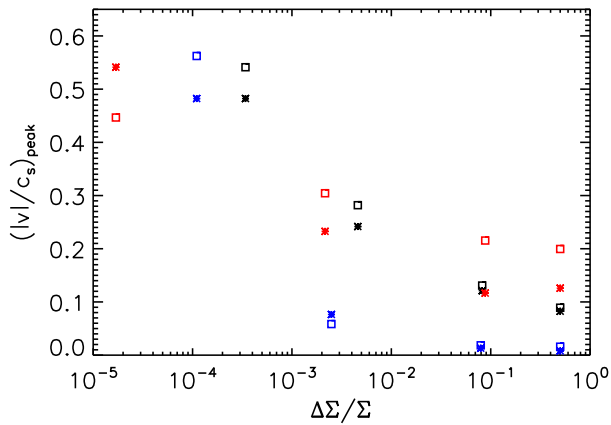}
\end{center}
\caption{
Peak of the turbulent velocity distribution versus fraction of the total surface density.  The squares are the planar velocity distribution peaks, and the asterisks are the
vertical velocity peaks.  The black symbols are the peak velocities for the MRI calculation at 50 AU (Resistive-50AU), blue is the MRI calculation at 4 AU (Resistive-4AU),
and red is the higher amplitude forced hydrodynamic turbulence run (Hydro-HA).  As one probes vertically deeper into the disk, the peak velocity decreases.
}
\label{vp_sigma}
\end{figure}

This paper represents a step toward making a direct connection between the simulated properties of turbulent protoplanetary disks and actual observations of these systems.
To this end, we have presented a series of calculations with varying physics from which we extracted the turbulent velocity distribution.   The simulations do not explicitly predict actual observables, as that would require the inclusion of radiation physics in one form or another.  However, our results do have several implications for the nature of disk turbulence in low mass protoplanetary disks that could potentially be tested with future observations, particularly those made with {\em ALMA}.

The first implication is that {\em if turbulence is driven solely by the MRI},
the turbulent linewidth ought to vary strongly as a function of both radius and height above the mid-plane. In regions of the 
disk where non-ideal effects are small, we predict mid-plane velocities that peak near 0.1 times the local sound speed. 
The characteristic turbulent velocities increase with height, such that significant regions of transonic flow occur in the 
atmosphere at $z > 3H$. This prediction is in at least rough agreement with the observational measurement of 
turbulent broadening in HD~163296, where \cite{hughes11} infer turbulent line widths consistent with a few tenths 
of the sound speed. The observational results for this system are then consistent with MRI-driven turbulence actually occurring 
in the outer disk. The upper limit for TW~Hya, on the other hand, is at best marginally consistent with the MRI prediction 
for near-ideal conditions -- pushing that limit lower has the potential to provide a stringent test of our models.

If we assume, on theoretical grounds, that the MRI is the only viable source of turbulence in low mass disks, then 
the agreement between our simulations and the observational results for HD~163296 is mildly encouraging. That is, MRI-driven turbulence produces the ``right" answer. This order of magnitude level of agreement, 
however, is nowhere near discriminating enough to exclude the possibility that some (unspecified) hydrodynamic 
instability is responsible for the observed turbulent broadening.  We have run purely hydrodynamic disk models that are driven to a turbulent state via large scale forcing, and these disks also show velocity fluctuations of the order of tenths of the sound speed, along with a velocity gradient between the mid-plane and the 
surface, where transonic velocities can be produced.

This raises the question, can observations distinguish between purely hydrodynamic turbulence and that driven by the MRI?  Of course, one option is to observe the magnetic field structure and amplitude (or lack thereof) in the disk itself, but this is currently exceedingly difficult \cite[but see][]{hughes09}.  If one cannot observe the field directly, the next best option is to observe a secondary effect of the magnetic field such as the turbulent velocity. However, as noted already, a single measurement of the turbulent velocity is certainly insufficient to distinguish between MHD and hydrodynamic drivers of turbulence. An arbitrary forcing of the fluid can still yield turbulent velocities that are more or less indistinguishable from those produced via the MRI.  

The prospects for distinguishing between different sources of turbulence are better if observations are able to probe either different heights within the disk (by exploiting multiple molecular species), different radii, or both. In a real disk, of course, there is no immediately accessible observable that isolates the turbulence at a particular physical height 
above the mid-plane. Rather, the observable is the degree of broadening of a given molecular line produced in a region of the 
disk where the temperature, density and chemistry yield sufficient emissivity, and the optical depth is not too high. Thus, where in $z$ a 
particular line is emitted will depend on the vertical structure of the disk itself.  We find that this structure differs significantly depending upon whether the disk is magnetized or not. Consider Fig.~\ref{dz_all}, which shows 
the time- and horizontally-averaged vertical density profile. The red curve is one of the forced hydro turbulence cases, and the other curves are a subset of the MRI-driven turbulence runs.  There is an obvious and significant difference between the density profiles at large $|z|$. The density departs significantly from Gaussian in the MRI cases.  The reason for this is that for $|z| \gtrsim 2H$, magnetic pressure dominates over gas pressure, and the gradient in magnetic pressure helps to support the gas against gravity. Thus, the gas pressure and density have a shallower slope in these regions. This magnetic and gas pressure structure is consistent with previous shearing box simulations \cite[e.g.,][]{hirose06}.

The difference in vertical structure between a magnetized and non-magnetized disk results in a distinct difference in the the profile of turbulent velocity with {\em column density}. This is shown in Figure~\ref{vp_sigma}, which plots how the characteristic turbulent velocity changes as a function of fractional column, defined as

\begin{equation}
\label{fractional_column}
\frac{\Delta \Sigma}{\Sigma} \equiv \frac{\int_z^{4H} \overline{\rho(z')} dz'}{\int_{-4H}^{4H}  \overline{\rho(z')} dz'}
\end{equation}

\noindent
where $\overline{\rho(z)}$ is the time- and horizontally averaged gas density (the time average is done from orbit 50 onwards).
There are two features to note in this plot.  The first is that, in general, the turbulent velocity decreases as one probes deeper into the disk.  This result was discussed above in the context of the velocity distributions, and is simply the result of velocity steepening in lower density regions.   The second feature is that the $v/\cs \sim 0.5$ values obtained in the upper disk regions can be found at a lower $\Delta\Sigma/\Sigma$ (by about an order of magnitude) in the hydro case versus the MHD cases.  This suggests that the column depth to which a particular line can probe may be very useful in determining the density structure away from the disk mid-plane and thus can constrain the turbulence mechanisms. 

Furthermore, the radial dependence of the velocity gradient with distance from the mid-plane may also be useful.  While the
general trend of decreasing $|v|/\cs$ with height is robust in all of our calculations, the presence of the MRI dead zone dramatically changes this gradient.
In particular, as Figure~\ref{vp_sigma} shows (blue points), the presence of a magnetically dead zone is quite obvious as the turbulent velocities drop well below $0.1\cs$ near the mid-plane region, but the active layers above and below the mid-plane, in combination with steepening, produce $|v|/\cs \sim 0.5$.  Thus, if one were to 
probe different depths into the disk and find a dramatic decrease in turbulent velocity towards the mid-plane, this would be strongly indicative of a dead zone region.

\section{Conclusions and Uncertainties}
\label{conclusions}

Our conclusions about the turbulent properties of low mass protoplanetary disks are:
\begin{enumerate}

\item Characteristic turbulent velocities are $\sim$ (0.1-1)$\cs$ for fully turbulent regions, in rough agreement with observations made to date \cite[e.g.,][]{hughes11}.  
These characteristic velocities are reasonably robust to variations in numerical (changes to local domain size) and physical (locations in a model disk) parameters.  

\item Turbulent velocity increases away from the disk mid-plane due to steepening.  In the upper region of the disk ($|z| \gtrsim 3 H$), the velocity distribution peaks around $0.5\cs$ and has
a significant ($\sim 10\%$) supersonic component.  As one probes towards the mid-plane, $|v|/\cs \sim 0.1$ is typical of fully turbulent disks. 

\item In calculations with an MRI dead zone near the mid-plane, the characteristic turbulent velocities are $\sim 0.01\cs$ within the dead zone.  In principle, with an improvement
in sensitivity, observations that probe different depths could see the presence of the dead zone as velocities drop from $\sim$ 0.1-1$\cs$ to $\sim 0.01\cs$.

\item The density structure for $|z| > 2H$ is significantly different in the MRI versus purely hydrodynamic cases, which could have potential implications for the observed
turbulent linewidths if different depths can be probed.

\item The vertical and planar velocity distributions are quite similar, suggesting that turbulent linewidths
will only weakly be dependent on the inclination angle.  

\end{enumerate}

Our predictions for the distribution of turbulent velocity in MRI-active disks suffer from a number of uncertainties. First, we have chosen to only focus on one non-ideal MHD effect, namely Ohmic resistivity.  Other non-ideal effects -- ambipolar diffusion and the Hall
term -- are also important in protoplanetary disks \citep{kunz04,bai11,wardle11}. Ambipolar diffusion, 
in particular, is important in low density regions, and may affect the properties of turbulence in the most 
observationally accessible location -- the disk atmosphere at large radius.
Moreover, the resistivity that we have employed neglects dust physics. We also caution that some of the most 
striking qualitative trends that we observe are linked to the steepening of waves near the disk surface. Wave 
propagation in disks is known to depend upon the vertical thermal structure \citep{bate02}, and hence the 
isothermal structure that we have assumed may not always be adequate. Even
before adding a treatment of the radiation physics, these limitations imply that there remains much work to be done to further constrain the turbulent velocities in simulations of MRI-active disks. 

While our results coupled with recent observations provide mild support for the model of MRI-driven angular momentum transport, the calculations that we have presented here have not been able to identify a strong observational discriminant between MRI-driven 
and purely hydrodynamic turbulence. Quite precise measurements of the turbulence as a function of 
height will be needed to tell one from the other on purely observational grounds.  In fact, if sufficiently high spatial resolution observations of the inner disk regions reveal
the presence of a dead zone region, then this would present very strong support for the MRI driving disk turbulence.

Finally, we reiterate that we have considered only arbitrary hydrodynamic forcing, rather than setting up known physical drivers of turbulence, such as 
self-gravity or even convection \citep{lesur10}.   By design, the average kinetic energy in the purely
hydrodynamic simulations nearly equals that of the MRI simulations.
If it could be established, theoretically, that hydrodynamic
drivers of turbulence were necessarily weaker than the MRI,
a single measurement of the turbulent velocity would then
distinguish between the two. If, on the other hand, we treat
the strength of hydrodynamic turbulence as a free parameter,
then our results suggest that the vertical variation of turbulent
velocities in the hydrodynamic and MHD limits can have a qualitatively similar trend. Of course, the physical mechanisms that might initiate hydrodynamic turbulence without arbitrary forcing could, in principle, imprint distinctive characteristics into the observable turbulent velocity field, which would allow them to be distinguished from the MRI 
more readily. To test this, it would be useful to repeat the analysis presented here for disks in which these other 
sources of turbulence are active. These calculations are currently underway and will be presented in future work.

\acknowledgments

We thank Meredith Hughes, Tilman Birnstiel, Charles Gammie, and John Hawley  for useful discussions and
suggestions regarding this work.  We also thank the anonymous referee whose comments greatly improved this paper. We acknowledge support from the NSF (AST-0807471, AST-0907872), from NASA's Origins of Solar Systems program (NNX09AB90G), from NASA's Astrophysics Theory program (NNX11AE12G), and
from the NSF through TeraGrid resources provided by the Texas Advanced Computing Center and the National Institute for Computational Science under grant number TG-AST090106.  We also acknowledge the Texas Advanced Computing Center at The University of Texas at Austin for providing HPC and visualization resources that have contributed to the research results reported within this paper. Computations were also performed on Kraken at the National Institute for Computational Sciences.

\end{document}